\documentclass[10pt,aps,showpacs,a4paper,floatfix,twocolumn,tightenlines]{revtex4}
\usepackage{epsfig}
\usepackage{psfig}
\usepackage{graphicx}
\usepackage{graphics}
\usepackage{xspace}
\usepackage{amssymb}
\usepackage{amsmath}
\usepackage{latexsym}
\usepackage{natbib}
\usepackage{mathrsfs}
\newcommand{\half}{\textstyle{1\over 2}}
\newcommand{\thalf}{\textstyle{3\over 2}}
\newcommand{\ds}{\displaystyle}

\begin{document}
\title{
Spectral function and relativistic mean-field description\\
of (un)polarized $(e,e'p)$ reactions: a consistent picture.
}

\author{Marco Radici and Andrea Meucci}
\affiliation{Dipartimento di Fisica Nucleare e Teorica, 
Universit\`{a} di Pavia, and\\
Istituto Nazionale di Fisica Nucleare, 
Sezione di Pavia, I-27100 Pavia, Italy}

\author{W.H. Dickhoff}
\affiliation{Department of Physics, Washington University, St. Louis, Missouri 63130, USA}

\date{\today}

\begin{abstract}
We analyze the unpolarized and polarized electron-induced proton-knockout
reactions on $^{16}$O in different kinematical settings
using two theoretical approaches. The first one is based on a relativistic
mean-field distorted-wave description of the bound and scattering states of
the proton, including a fully relativistic electromagnetic current operator. 
The second approach adopts the same current operator, but describes the
proton properties consistently on the basis of microscopic calculations of the
self-energy in ${}^{16}$O below the Fermi energy and final-state damping in
nuclear matter above the Fermi energy, using the same realistic short-range
and tensor correlations.
Good agreement with all unpolarized and polarized data is obtained at low and
high $Q^2$ by using the same spectroscopic factors fixed by the low-$Q^2$
analysis, indicating that a high degree of internal consistency has been
reached.
\end{abstract}


\maketitle

A long series of high-precision experiments on several
nuclei~\cite{frou,mou,bern,dewitt,lapik} have generated  
a well established tradition which singles out exclusive $(e,e'p)$ knockout
reactions as the primary tool to explore the single-particle aspects of the
nucleus.
The experimental analysis has focused on the missing energy spectrum of the
nuclear response, assigning specific quantum numbers 
and spectroscopic factors to the various peaks corresponding to orbitals 
close to the Fermi energy.
In addition, the missing momentum dependence of these spectra has been
studied, stimulating for example the exploration of the high-momentum
components induced by nucleon-nucleon correlations inside 
nuclei~\cite{mudi,mupodick}. 
The theoretical description of these reactions have usually been performed
in the framework of the nonrelativistic distorted-wave impulse approximation
(DWIA), including the Coulomb distortion of the electron and 
proton waves due the presence of the nuclear
field~\cite{frou,bgprep,book,kellyrep}. 
This approach was able to describe to a high degree 
of accuracy the shape of the experimental momentum distribution for
several nuclei in a wide range of different kinematics~\cite{book,kellyrep}.
However, a systematic rescaling of the normalization of the 
bound state, interpreted as the spectroscopic factor for the corresponding
level, had to be applied in order to reproduce the magnitude of the experimental
distribution~\cite{lapik,pandharep}. This systematic
deviation from the mean-field expectations has clearly identified the limits
of this approximation.
In fact, nowadays a clear picture has emerged in which
a considerable mixing between single-hole states and more complicated
configurations results in a fragmentation of the
single-particle strength in several peaks around and beyond the Fermi surface.
A further depletion of the single-particle strength is induced by short-range
and tensor correlations between nucleon pairs in the 
ground state~\cite{dimu,dickrep}. 

More recent $(e,e'p)$ experiments have been carried out at the Jefferson
Laboratory (JLAB)~\cite{e89003,e89033} at higher momentum transfer $Q^2$ and
with increased statistics such that the fully differential cross section
is now directly available. The new kinematic domain required a 
substantial upgrade of several theoretical ingredients in order to incorporate
all possible relativistic effects. Models based on relativistic DWIA (RDWIA)
have been developed, where the Dirac equation is solved directly 
for the nucleon bound and scattering states~\cite{pickvanord,jinonl,spain,heda}
or, equivalently, a Schr\"odinger-like equation is solved 
and the spinor distortion by the Dirac scalar and vector potentials is
incorporated in an effective 
current operator in the so-called effective Pauli reduction~\cite{heda,kelly}.
A successful description of the data has been achieved, but slightly different
spectroscopic factors are deduced, because the 
relativistic optical potentials in general give a stronger residual 
final-state interaction (FSI) than 
the corresponding nonrelativistic ones~\cite{spain,bgpc}.
Moreover, the limits of validity of the older 
DWIA analysis versus RDWIA were not always properly explored, as discussed in
Ref.~\cite{mgp}, resulting, for example, in a certain degree of ambiguity for
the spectroscopic factors extracted at low energy.

Despite several sources of theoretical uncertainties (different equivalent
potentials for FSI, relativistic effects on both FSI and spectroscopic factors,
off-shell effects...), a consistent microscopic treatment of the $(e,e'p)$
reaction mechanism at different kinematics is highly desirable. 
Results for a first attempt towards this goal were recently obtained
in Ref.~\cite{rdr} (see also Ref.~\cite{ryck} concerning the treatment of FSI), 
where a successful analysis of low- and high-$Q^2$ data was performed using 
identical spectroscopic factors which were deduced at low $Q^2$. 
In the present paper, this analysis is extended to the corresponding JLAB 
experiment with polarization, as reported in Ref.~\cite{e89033}.
The results will then be compared with those obtained in the RDWIA approach of
Ref.~\cite{mgp}, where a consistent description of low- and high-energy data
was generated and a careful analysis of the limits of the nonrelativistic DWIA
was carried out. The sensitivity to different off-shell prescriptions
for the electromagnetic current operator will be also discussed~\cite{meu}, 
but the difference between spectroscopic factors obtained by nonrelativistic 
and relativistic analyses remains unsolved and its discussion is beyond 
the scope of this paper.

The basic ingredient of the calculation is the transition amplitude 
(omitting spin degrees of freedom for simplicity)~\cite{book,rdr}
\begin{eqnarray}
J^{\mu}_n (\omega ,\vec q, \vec p_N^{\, \prime}, E_R) = \int d \vec p 
\  d \vec p^{\, \prime} \  
\chi^{\left( -\right)\, *}_{p'_{\scriptscriptstyle N} 
E_{\scriptscriptstyle R} n} (\vec p^{\, \prime}) \nonumber \\  
\times \  {\hat J}_{\rm eff}^{\mu}
(\vec p, \vec p^{\, \prime}, \vec q, \omega) \  
\phi^{}_{E_{\scriptscriptstyle R} n} (\vec p) \  [Z_n(E_R)]^{1\over 2} \  , 
\label{eq:scattampl}
\end{eqnarray}
where $\vec q, \omega$ are the momentum and energy transferred to the target
($Q^2 = q^2 - \omega^2$) and $\vec p_N^{\, \prime}$ is the knocked-out nucleon
momentum, leaving the residual nucleus in a well-defined state with energy 
$E_R$ and quantum numbers $n$. The function $\phi^{}_{E_{\scriptscriptstyle R} n}$ 
describes the overlap between the exact $A-$body initial state
and the residual $(A-1)-$body state induced by producing a hole; 
$\chi^{\left( -\right)}_{p'_{\scriptscriptstyle N} E_{\scriptscriptstyle R} n}$
describes the same kind of overlap when producing the hole in the exact
$A$-body final state~\cite{book}. The norm 
of $\phi^{}_{E_{\scriptscriptstyle R} n}$ is 1 and $Z_n (E_R)$ is the
spectroscopic factor associated with the removal process, i.e., it corresponds
to the probability that the residual nucleus can indeed be considered 
as the target nucleus with a hole. 
The boundary conditions of the eigenvalue problem for 
$\chi^{\left( -\right)}_{p'_{\scriptscriptstyle N} E_{\scriptscriptstyle R} n}$
are those of an incoming wave. 

In the RDWIA of Refs.~\cite{mgp,meu}, $\phi^{}_{E_{\scriptscriptstyle R} n}$ 
is replaced by the solution of a Dirac equation~\cite{adfx} deduced in the
context of a relativistic mean-field theory 
that satisfactorily reproduces global and single-particle properties of several
nuclei~\cite{lala}. 
For the scattering states the effective Pauli reduction is applied.
The Darwin nonlocality factor, that contains the effect
of the negative-energy components of the spinor, is reabsorbed in the current
operator, which becomes
an effective relativistic one-body operator depending on the Dirac scalar
and vector potentials~\cite{heda,kelly}, as well as on the chosen
off-shell prescription (cc1, cc2, or cc3)~\cite{defo,meu}. 
The function $\chi^{\left( -\right)}_{p'_{\scriptscriptstyle N}
E_{\scriptscriptstyle R} n} \sim 
\chi^{\left( -\right)}_{p'_{\scriptscriptstyle N}}$ becomes a two-component
spinor which solves the
Schr\"odinger equation with the equivalent central and spin-orbit potentials 
expressed in terms of the original Dirac scalar and vector 
ones~\cite{cooper}. 

In Ref.~\cite{rdr} the transition amplitude is evaluated by systematically
applying the effective Pauli reduction to both the initial and final Dirac
spinors, determining the relevant integrals in momentum space
thus avoiding any effective momentum approximation (EMA)~\cite{kelly}.
The current operator displays the same features as in the RDWIA
discussed above, i.e., it is an effective one-body relativistic operator 
depending on the Dirac scalar and vector potentials. In Ref.~\cite{rdr}
only the cc1 off-shell prescription has been
considered for compatibility with older low-energy data analyses~\cite{prbdm}. 
The scattering state of the (very energetic) proton is described in the eikonal
approximation by a uniformly damped plane wave, or, equivalently, by a 
plane wave with a complex momentum 
$\vec p_f^{\, \prime} = \vec p_N^{\, \prime} + i \vec
p^{}_I$~\cite{br,cannata}. The imaginary part $p_I$ is microscopically
justified by linking the proton absorption to the same process taking place
in nuclear matter and by calculating the nucleon self-energy in a
self-consistent manner with realistic short-range and tensor
correlations~\cite{dr,rdr}. 
The observed damping is also in agreement with experimental expectations in
different kinematic domains~\cite{ne18}, however, embedding the proton in
nuclear matter prevents the inclusion of spin-orbit effects; therefore,
the corresponding Darwin nonlocality factor for the final state is just 
1. The function $\phi^{}_{E_{\scriptscriptstyle R} n}$ is obtained from
$p$-shell quasihole states deduced from the nucleon self-energy calculated
for ${}^{16}$O using realistic short-range and tensor 
correlations~\cite{mupodick}. 

The hadronic tensor of the reaction, $W^{\mu \nu}$, involves an average over
initial states and a sum over the undetected final states of bilinear products 
of the scattering amplitude (\ref{eq:scattampl}). 
The differential cross section for the $(\vec e, e' \vec p)$ reaction, with 
initial beam helicity $h$ and proton polarization component $\hat s$, 
becomes~\cite{book}
\begin{equation}
\begin{array}{l}
{\ds {{d \sigma_{h \, \hat s}} \over {d \vec p_e^{\, \prime}
d \vec p_N^{\, \prime} } }  = { e^4 \over {16 \pi^2}} {1 \over Q^4 p_e p'_e }}\
L^{}_{\mu \nu} W^{\mu \nu} \\[.5cm]
\equiv {\ds { e^4 \over {16 \pi^2}} {1 \over Q^4 p_e p'_e } }\
L^{}_{\mu \nu} \  \ 
{\lower7pt\hbox{$_i$}} \kern-7pt \hbox{$\overline \sum$} \  
\hbox{\hbox {$\sum$} \kern-15pt {$\displaystyle \int_f$\ } } J^{\mu}_n
J^{\nu \, *}_n  \  \delta 
\left( E^{}_f - E^{}_i - \omega \right) \\[.5cm]
= {\ds {{d \sigma^o} \over {d \vec p_e^{\, \prime} d \vec p_N^{\, \prime} }}
\  \frac{1}{2} }  \   
\left[ 1 + \vec P \cdot {\hat s} + 
h \left( A + \vec P^{\, \prime} \cdot {\hat s} \right) \right]  \  ,
\end{array} 
\label{eq:cross}
\end{equation}
where $p^{}_e, p'_e$ are the initial and final electron momenta and
$L_{\mu \nu}$ is the lepton tensor. 
The coefficients of the linear expansion are the induced polarization 
$\vec P$, the electron analyzing 
power $A$, and the polarization transfer coefficient $\vec P^{\, \prime}$.
The reference frame in the polarimeter is formed by the direction of
$\vec p_N^{\, \prime}$ ($L$ component), the direction of 
$\vec q \times \vec p_N^{\, \prime}$ ($N$ component) and $\hat N \times
\hat L$ ($T$ component). In coplanar kinematics, as is the case for the 
E89033 experiment at JLAB~\cite{e89033}, only 
$P^N, P^{\prime \, L}$ and $P^{\prime \, T}$ survive.
When summing over the recoil proton polarization and the beam helicity,
the usual unpolarized cross section $d \sigma^o$ is recovered. 


\begin{figure}[th]
\includegraphics[height=9.6cm, width=8.4cm]{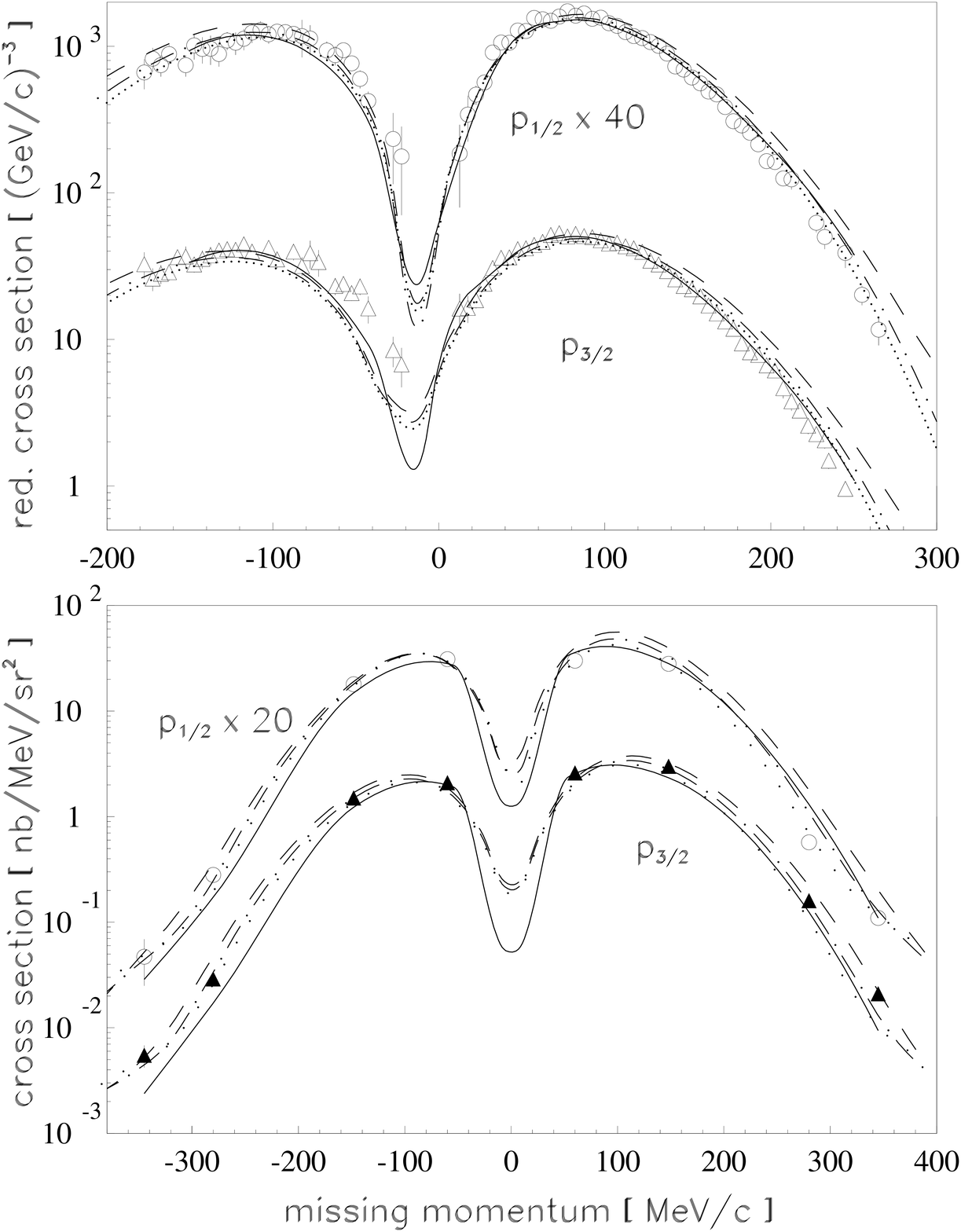} 
\caption {Upper panel: cross section for the $^{16}$O$(e,e'p)^{15}$N
reaction at $E_p = 90$ MeV constant proton energy in the center-of-mass 
system in parallel kinematics~\protect\cite{nikhef}. Lower panel: 
cross section for the same reaction but at $Q^2=0.8$ (GeV/$c)^2$ in 
perpendicular kinematics~\protect\cite{e89003}. Data 
for the $p\half$ state have been multiplied by 40 and 20, respectively.
Solid lines show the results when using the quasi-hole spectral function
for the bound state (see text) with spectroscopic factors $Z_{p1/2}=0.644$ 
and $Z_{p3/2}=0.537$ in both panels~\protect\cite{prbdm,rdr}. 
Dashed, dot-dashed, and dotted lines 
represent the result of the RDWIA approach with cc1, cc2, cc3 off-shell 
prescriptions, respectively (see text). All the RDWIA curves in both panels 
have been rescaled by the spectroscopic factors $Z_{p1/2}=0.708$ and  
$Z_{p3/2}=0.602$, obtained by a $\chi^2$ fit to the data of 
Ref.~\protect\cite{nikhef} using the cc3 current.}
\label{fig:fig1}
\end{figure}


In Fig.~\ref{fig:fig1} we first reconsider the unpolarized $^{16}$O$(e,e'p)$
reaction leading to the ground state and the first excited state of $^{15}$N
with $p\half$ and $p\thalf$ quantum numbers,
respectively. In the upper panel, data have been collected in parallel
kinematics $(\vec p_N^{\, \prime} \parallel \vec q)$ at a constant proton
energy of 90 MeV in the center-of-mass system~\cite{nikhef}.
They are presented in the form of the reduced cross section 
\begin{equation}
n(\vec p_m, E_m) \equiv \frac{d \sigma^o}{d \vec p_e^{\, \prime} 
d \vec p_N^{\, \prime}} \  \frac{1}{K \sigma_{ep}} \ ,
\label{eq:redxsect}
\end{equation}
as a function of the missing momentum
$\vec p_m = \vec p_N^{\, \prime} - \vec q$ at the considered
missing energy $E_m$, where $K$ is a suitable kinematic factor and
$\sigma_{ep}$ is the elementary (half off-shell) electron-proton cross
section~\cite{defo}. For ease of viewing, the results for the 
transition to the $p\half$ ground state have been multiplied by 40.
The solid lines refer to the calculations employing the $p$-shell quasihole 
states for $^{16}$O in a nonrelativistic framework, as 
discussed in Ref.~\cite{prbdm}. The spectroscopic factors extracted from the
data are $Z_{p1/2}=0.644$ and $Z_{p3/2}=0.537$, respectively.
The dashed lines show the results of the RDWIA analysis 
with the same cc1 off-shell prescription; dot-dashed and
dotted lines indicate the results when using the cc2 and cc3 recipes,
respectively. Hence, the comparison among dashed, 
dot-dashed, and dotted lines shows the evolution in $p_m$ of the theoretical
uncertainty related to offshellness at this kinematics. Relativistic 
mean-field bound states are obtained by solving Hartree-Bogoliubov 
equations with finite-range interactions~\cite{adfx}. The proton 
scattering wave is deduced from relativistically equivalent energy-dependent optical 
potentials~\cite{cooper}. The resulting spectroscopic factors, $Z_{p1/2}= 0.708$ and 
$Z_{p3/2}=0.602$, have been obtained by a $\chi^2$ fit using the cc3 current, which 
gives an overall better description of the $(e,e'p)$ observables, particularly for the 
left-right asymmetry. In the lower panel, the same reaction is considered
at constant $(\vec q, \omega)$ with $Q^2 = 0.8$ 
(GeV/$c$)$^2$~\cite{e89003}. The data now refer to the fully differential
unpolarized cross section $d\sigma^o$, avoiding any ambiguity in modeling the
off-shell behavior of $\sigma_{ep}$~\cite{defo}. 
Again the solid lines refer to the calculation employing the $p$-shell
quasihole states but with an effective relativistic current operator and an
eikonal microscopic description of FSI as discussed 
above (see Ref.~\cite{rdr} for further details). 
The dashed, dot-dashed, and dotted lines still refer to the RDWIA analysis 
with the cc1, cc2, and cc3 off-shell prescriptions
for the electromagnetic current, respectively. 
The $p\half$ results are multiplied by a factor 20. The 
theoretical curves are rescaled by the same spectroscopic factors as in the 
upper panel. The agreement with the data remains very good also in this case.
This confirms the internal consistency of the two approaches, since the
spectroscopic factors correspond to a nuclear property that must be 
independent of the probe scale $Q^2$. Incidentally, we remark that an extraction 
of spectroscopic factors directly from the data of Ref.~\cite{e89003} most 
likely produces ambiguous results, due to the small number of data points (8 only) of 
the experiment. We tried such an extraction using the RDWIA curves, 
but the fits had very high $\chi^2$ per degree of freedom and gave $p\thalf$ rescaling 
coefficients which are systematically bigger than the $p\half$ ones, contrary to any 
reasonable expectation (see also Ref.~\cite{e89003}). The 
approximation introduced in the eikonal treatment of FSI, specifically the
absence of any spin-orbit effect, does not affect the agreement between the
solid lines and the data. Similarly, the sensitivity to the off-shell
ambiguity in the electromagnetic current operator is relatively weak. 
After all, it is well known that the cross section is not particularly
sensitive to the theoretical uncertainties in the description of FSI
and the off-shell prescriptions within a range of about 10\%.
In Ref.~\cite{rdr} more sensitive observables in the unpolarized cross section
were considered. Good agreement with these data was maintained but, at the same
time, the limitations of this approximation emerged, 
particularly in the left-right asymmetry.
Here, for the same kinematic conditions we extend the 
analysis to polarization observables.


\begin{figure}[h]
\includegraphics[height=7.6cm, width=8.6cm]{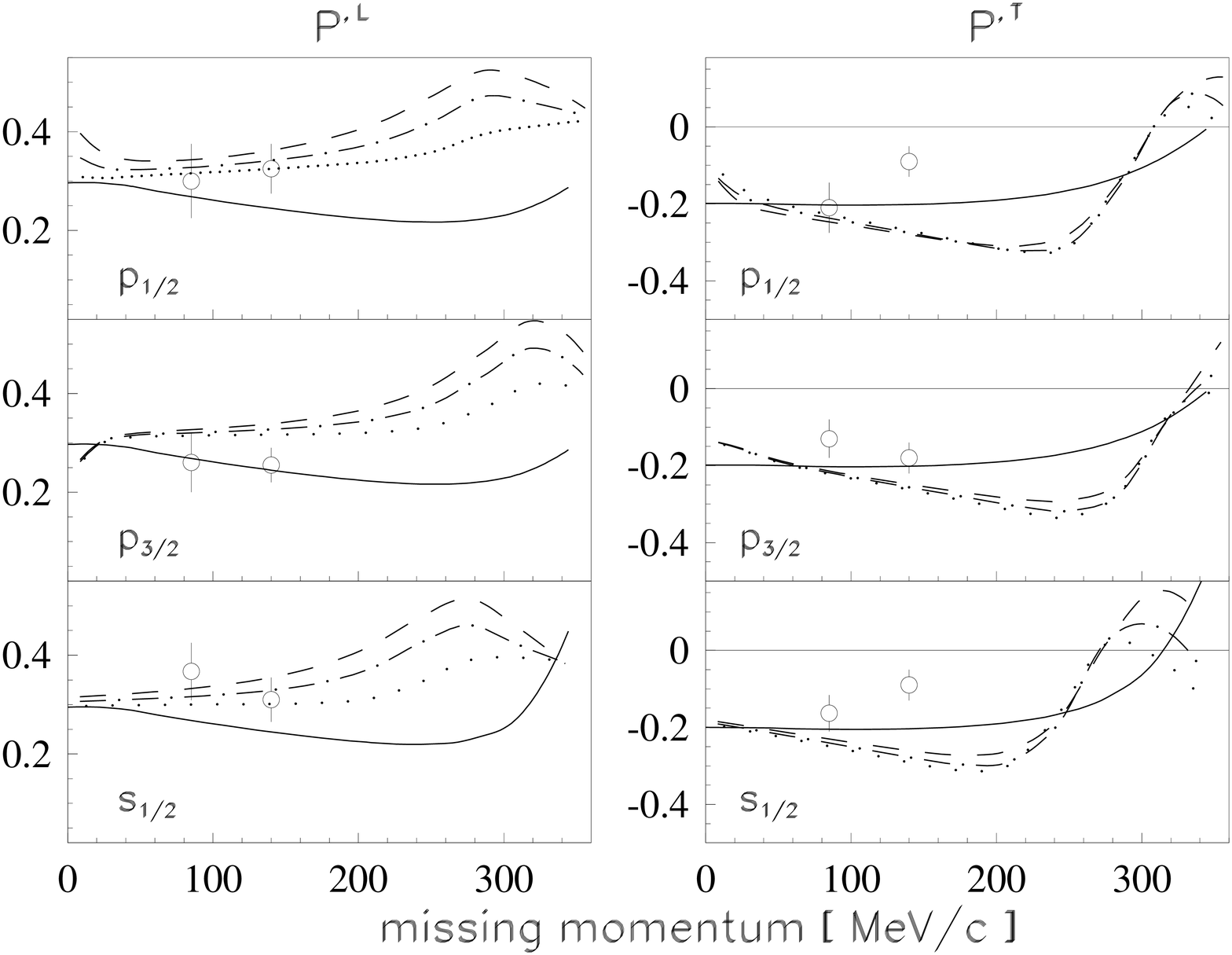} 
\caption {Polarization transfer components $P^{\prime \, L}, P^{\prime \, T}$ 
for the $^{16}$O($\vec e,e'\vec p$) reaction at 
$Q^2=0.8$ (GeV/$c)^2$ in perpendicular kinematics~\protect\cite{e89033}
leading to the 
$^{15}$N $p\half$, $p\thalf$ and $s\half$ residual states. 
Solid, dashed, dot-dashed, and dotted lines as in Fig.~\protect\ref{fig:fig1}.}
\label{fig:fig2}
\end{figure}


\begin{figure}[th]
\includegraphics[height=6.8cm, width=8.3cm]{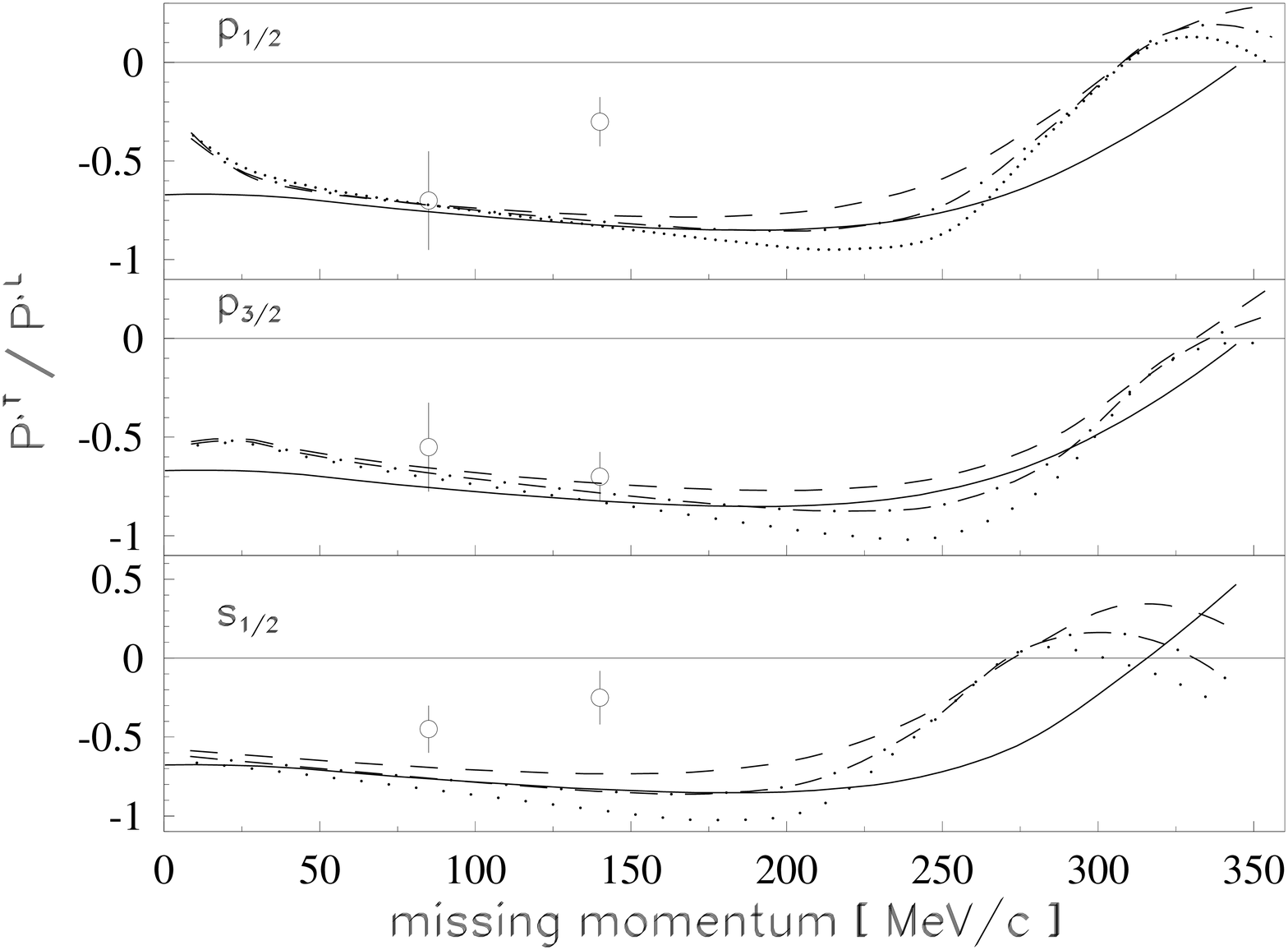}
\caption  {The ratio $P^{\prime \, T} / P^{\prime \, L}$ 
for the $^{16}$O($\vec e,e'\vec p$) reaction at 
$Q^2=0.8$ (GeV/$c)^2$ in perpendicular kinematics~\protect\cite{e89033}
leading to the 
$^{15}$N $p\half$, $p\thalf$ and $s\half$ residual states. Notations as in 
Fig.~\protect\ref{fig:fig1}.}
\label{fig:fig3}
\end{figure}


In Figs.~\ref{fig:fig2} and \ref{fig:fig3} the polarization transfer
components $P^{\prime \, L}, P^{\prime \, T}$ and their ratio
$P^{\prime \, T}/P^{\prime \, L}$ are shown as 
functions of the missing momentum $p_m$, respectively, for 
the $^{16}$O($\vec e,e'\vec p$) reaction at $Q^2=0.8$ (GeV/$c)^2$ and
constant $(\vec q, \omega)$ for 
the transitions to the $^{15}$N ground state $p\half$, the first $p\thalf$ 
state at $E_m=6.32$ MeV and the weak peak with quantum numbers
$s\half$ rising above a continuum background at
$E_m \sim 28$ MeV~\cite{e89033}. 
Solid, dashed, dot-dashed, and dotted lines refer to the same 
calculations as in Fig.~\ref{fig:fig1}. For these observables and at this
kinematics, the sensitivity to off-shell effects is at most $\lesssim 15$\%.  
The overall agreement with the data is still good, 
particularly for the microscopic calculations with the $p\thalf$ quasihole
state that performs even better than the RDWIA analysis presented here or
obtained by other groups~\cite{e89033}. This fact is 
remarkable, since the RDWIA analysis depends on mean-field phenomenological 
potentials with several parameters fitted to the considered target
and energy domain, while the calculation with 
quasihole states is basically parameter free.
In fact, from Eq.~(\ref{eq:cross}) it is easy to verify 
that the polarization observables are given by ratios between a specific spin
projection of the cross section and the unpolarized cross section, eliminating
any sensitivity to the spectroscopic factor 
which is anyway fixed from the very beginning to the low-energy data of
Ref.~\cite{nikhef}. Moreover, the calculations of the solid lines include an
attempt of a microscopic description of FSI in the framework of the eikonal
approximation in a way which is consistent with the description of the bound
state. 
The limitations of such an approach are more evident in the $j=\half$ case,
where, contrary to the RDWIA analysis, the absence of any spin-orbit 
effects is most likely responsible for the worse agreement. In any case, the
second $P^{\prime \, T}$ data point for both $p\half$ and $s\half$ shells 
appears not reproducible in both calculations, causing the 
theoretical ratio $P^{\prime \, T}/P^{\prime \, L}$ to deviate 
substantially from the experiment. 

In summary, we have analyzed the unpolarized and polarized proton knockout
reactions on $^{16}$O at
different kinematics with two theoretical approaches. The RDWIA is based on a
relativistic mean-field description of the proton bound state and on the
effective Pauli reduction of the final Dirac spinor,
leading to a Schr\"odinger-equivalent mean-field description of residual FSI
and to an effective relativistic electromagnetic current operator which
depends on the Dirac scalar and vector potentials. The 
same kind of Pauli reduction (and resulting current operator) is used for both
initial and final states in the second approach, where a microscopic 
description of the bound state properties is obtained 
by solving the Dyson equation with a nucleon self-energy which includes
realistic short-range and tensor correlations for $^{16}$O. 
As an attempt towards full consistency, the proton scattering wave is then
generated in the eikonal approximation by microscopically
calculating the damping of a plane wave as a solution of the Dyson equation
for the nucleon self-energy including the same realistic short-range and tensor
correlations between the struck proton and the
surrounding nucleons in nuclear matter. A systematic good agreement with data
is observed for both unpolarized and polarized reactions at low and high
$Q^2$ by using the same spectroscopic factors fixed
by the low-$Q^2$ analysis, thus indicating that a high degree of internal
consistency has been reached.

\vspace{.5truecm}

This work is supported by the U.S. National Science Foundation under Grant No. 
PHY-9900713. We acknowledge fruitful discussions with C. Giusti and F.D. Pacati.

\end{document}